\documentstyle[preprint,eqsecnum,cite,epsf,aps]{revtex}
\begin{document}

\title{\Large A Consistent Noncommutative  Field Theory: the
  Wess-Zumino Model}

\author{H. O. Girotti$^{\,a}$, M. Gomes$^{\,b}$,
  V. O. Rivelles$^{\,b}$ and A. J. da Silva$^{\,b}$}  
\address{$^{a\,}$Instituto de F\'\i sica, Universidade Federal do Rio Grande
do Sul\\ Caixa Postal 15051, 91501-970 - Porto Alegre, RS, Brazil\\ E-mail:
hgirotti@if.ufrgs.br}
\address{$^{b\,}$Instituto de F\'\i sica, Universidade de S\~ao Paulo\\
 Caixa Postal 66318, 05315-970, S\~ao Paulo - SP, Brazil\\
E-mail: mgomes, rivelles, ajsilva@fma.if.usp.br}

\maketitle

\begin{abstract}
We show that the noncommutative Wess-Zumino model is
renormalizable to all orders of perturbation theory. The
noncommutative scalar potential by itself is 
non-renormalizable but the Yukawa terms demanded by supersymmetry
improve the situation turning the theory into a renormalizable
one. As in the commutative case, there are neither quadratic nor linear
divergences. Hence, the IR/UV mixing does not give rise to
quadratic infrared poles. 
\end{abstract}

\section{INTRODUCTION}
Noncommutative geometry has been receiving a great deal of attention
in the context of string/M-theory. Initially it appeared as a possible
compactification manifold of space-time \cite{Connes} and led to the
appearance of noncommutative quantum field theories on noncommutative
tori \cite{Morariu,Hofman,Krajewski,Sheikh}. More recently
\cite{Seiberg-Witten} it was shown that the dynamics of a D-brane in
the presence of a $B$-field can, in certain limits, be described by a
deformed gauge field theory in terms of Moyal products on space-time.
Since this field theory arose from a coherent truncation of a string
theory it is expected that deformed field theories are consistent by
themselves.  This motivated an intensive investigation of
noncommutative quantum field theories on four dimensional Euclidean
and Minkowski spaces. Scalar fields
\cite{Chepelev,Minwalla,Fischler,Arefeva,Arefeva-complex}, gauge
fields
\cite{Martin,Hayakawa,Grosse,Madore,Hayakawa2,Benaoum,Sheikh-Jabbari2,Matusis,Gracia,Bonora,Arefeva-last,Bilal,Rajaraman,Armoni}
and supersymmetric theories \cite{Chu,Ferrara,Terashima} have been
studied. Some two dimensional models have also been analyzed
\cite{Grosse2,Moreno,Furuta,Schiappa}.

A distinct characteristic of a class of noncommutative quantum field
 theories is the mixing of ultraviolet (UV) and infrared (IR)
 divergences \cite{Minwalla} reminiscent of the UV/IR connection of
 string theory. For the $\phi^4_4$ massive scalar field there is an
 infrared quadratic singularity  in the propagator at the one loop level, which
jeopardizes the perturbative formulation of the theory.
On the other hand, the theory has been proved to remain ultraviolet
renormalizable up two loops \cite{Arefeva} although this 
does not seem to hold at all orders \cite{Chepelev}. Also,
models involving complex scalar fields are not always renormalizable
not even at one-loop approximation \cite{Arefeva-complex}.
Therefore, it is relevant to
understand the renormalizability properties of noncommutative field
theories to find out whether they are consistent. 

It has been suggested that, due to the absence
of quadratic divergences in their commutative version,  noncommutative
supersymmetric theories  may remain ultraviolet renormalizable
\cite{Chepelev,Ferrara}. 
The 
superspace formulation has already been accomplished at the classical
level \cite{Ferrara,Terashima}. However, at the quantum level only
one loop results have been reported for supersymmetric gauge
theories. As in the commutative case only logarithmic divergences
show up \cite{Matusis,Arefeva-last}.

This paper is dedicated to show that the noncommutative Wess-Zumino
model in four dimensions is a consistent quantum field theory in the
sense of being ultraviolet renormalizable and free of the IR/UV mixing
at any arbitrary order of perturbation. This happens even though the
scalar potential of the noncommutative Wess-Zumino model belongs to
the class of non-renormalizable theories discussed in
\cite{Arefeva-complex}. It is a potential typical of a F-term $\phi^*
\star \phi^* \star \phi \star \phi$ (while a D-term induces $\phi^*
\star \phi \star \phi^* \star \phi$) but nevertheless supersymmetry
still eliminates all quadratic divergences. This is at the root of the
renormalizability of the model.

Noncommutative field theories containing just scalar and fermion
fields, as is the case in the Wess-Zumino model, are constructed from the usual
Lagrangian by replacing the ordinary product by the Moyal product of
fields, i. e.  $ A B \rightarrow A \star B$. The Moyal product is
noncommutative and obeys the rule

\begin{eqnarray}
&&\int dx \phi_1(x)\star\phi_2(x)\star...\star\phi_n(x)=\nonumber \\
&& \int \prod \frac{d^4k_i}{(2\pi)^4} (2\pi)^4\delta(k_1+k_2+\ldots +
k_n)\tilde\phi_1(k_1) \tilde \phi_2(k_2)\ldots \tilde\phi_n(k_n)
\exp(i\sum_{i<j}k_i\wedge k_j), \label{1}
\end{eqnarray}

\noindent
where $\tilde \phi_i$ is the Fourier transform of the field $\phi_i$,
the index $i$ being used to distinguish different fields. In (\ref{1})
we have introduced the notation $a\wedge b= 1/2 a^\mu b^\nu \Theta_{\mu\nu}$, 
where $\Theta_{\mu\nu}$ is the anti-symmetric constant matrix characterizing
the noncommutativity of the underlying space. We shall assume from now on
that $\Theta_{0i}=0$ in order to evade causality and unitarity problems 
\cite{Toumbas}.

To represent Feynman amplitudes one could either use a double line
notation, as the one introduced by 't Hooft for matrix models, or
single lines, which demands the symmetrization of the kernel (\ref{1})
over the arguments of fields of the same kind. In this work we adopt
the second systematics. 

The paper is organized as follows. In Section II we present and
discuss general aspects of the noncommutative Wess-Zumino model. The
one loop analysis is performed in Section III, while in Section IV we
demonstrate the renormalizability of the model to all orders of
perturbation theory. Section V contains some final comments and the
conclusions. 

\section{THE NONCOMMUTATIVE WESS-ZUMINO MODEL}

In four dimensional Minkowski space-time the Wess-Zumino model is defined by the Lagrangian density \cite{Wess}

\begin{eqnarray}
{\cal L} &= &\frac12  A(-\partial^2)A+\frac12  B(-\partial^2)B
+ \frac12 \overline \psi(i\not \! \partial - m)\psi +\frac12 F^2+\frac12 G^2
+ m F A + m G B +\nonumber \\
&\phantom a & g (F A^2 -  F B^2 + 2 G A B -  \overline \psi
\psi A - i  \overline \psi \gamma_5\psi B ), \label{2}
\end{eqnarray}

\noindent
where $A$ is a scalar field, $B$ is a pseudo scalar field, $\psi$ is
a Majorana spinor field and $F$ and $G$ are, respectively, scalar and
pseudoscalar auxiliary fields. By extending the above model to a
noncommutative space one is led to the Lagrangian density

\begin{eqnarray}
{\cal L} &=&\frac12  A(-\partial^2)A+\frac12  B(-\partial^2)B
+ \frac12 \overline \psi(i\not \! \partial - m)\psi +\frac12 F^2+\frac12 G^2
+ m F A + m G B + \nonumber
\\ &\phantom a& g (F\star A \star
A- F\star B \star B + G\star A \star B + G \star B \star A -\overline \psi
\star \psi\star A - \overline \psi\star  i\gamma_5 \psi \star B). \label{3}
\end{eqnarray}

 It should be noticed that there is only one
possible extension of the cubic term $2 G A B$, to the noncommutative
case, which preserves supersymmetry. It should also be emphasized that
the noncommutative supersymmetry transformations are identical to the
commutative ones since they are linear in the fields and no Moyal
products are, therefore, involved. Hence, the extension of the theory
to the noncommutative case does not alter the form of the Ward
identities, which in turn implies that all fields have vanishing
vacuum expectation values. 

The elimination of the auxiliary fields through their corresponding
equations of motion turns the bilinear terms in the Lagrangian
Eq.(\ref{3}) 
into the standard mass terms. On the other hand, the cubic terms
produce quartic interactions which, in terms of a complex field
$\phi=A+iB$, can be cast as $\phi^* \star \phi^* \star \phi \star
\phi$. This potential belongs to a class of non-renormalizable
potentials, as discussed in \cite{Arefeva-complex}. As it will be
shown below, supersymmetry saves the day turning the theory into a
renormalizable one. 

The Lagrangian (\ref{3}) was also written using the superspace formalism in
\cite{Ferrara,Terashima}. However, we will work with components fields in
order to trace the effects of noncommutativity in the divergent 
Feynman integrals.

\begin{figure}
\centerline{\epsfbox{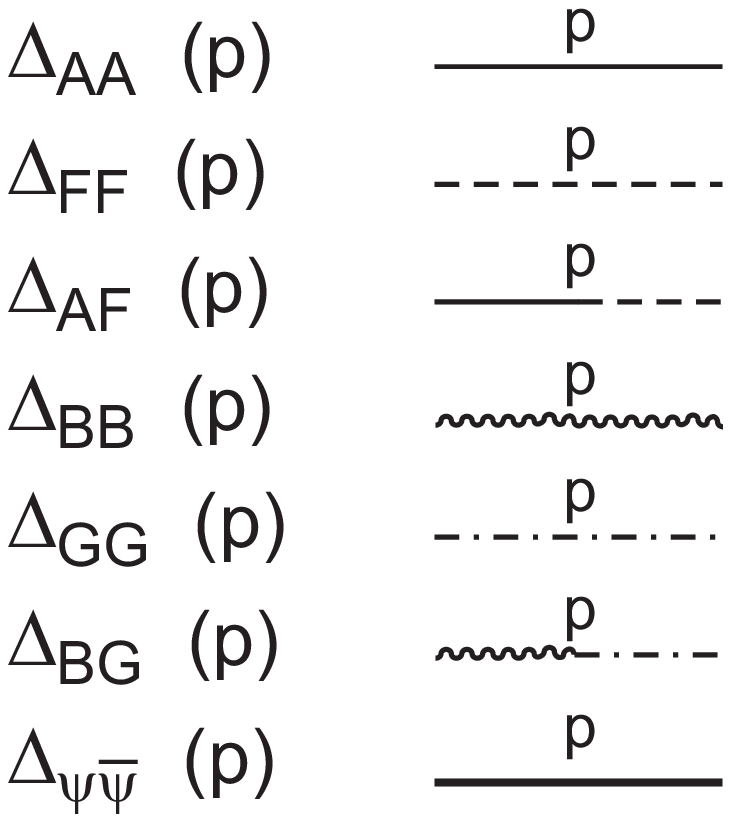}}
\caption{{\it Graphical representation for the propagators.}}\label{Fig1}
\end{figure}

The propagators for the $A$ and $F$ fields are (see Fig. \ref{Fig1})

\begin{mathletters}
\begin{eqnarray}
\Delta_{AA}(p) &=& \Delta(p)\equiv\frac{i}{p^2-m^2+i\epsilon},\\
\Delta_{FF}(p) &=& p^2 \Delta(p),\\
\Delta_{AF}(p) &=&\Delta_{FA}(p) = -m \Delta(p),
\end{eqnarray}
\end{mathletters}

\noindent
whereas the propagators involving the $B$ and $G$ fields have identical
expression (i.e., they are obtained by replacing $A$ by $B$ and $F$ by $G$).
For the $\psi$ field we have

\begin{equation}
S(p)= \frac{i}{\not \! p -m}.\label{4}
\end{equation}

The analytical expressions associated to the vertices are:

\begin{mathletters}
\begin{eqnarray}\label{4a}
F A^2 \quad {\mbox {vextex:}}&& \quad ig \cos(p_1\wedge p_2), \\
F B^2  \quad {\mbox {vextex:}}&& \quad -ig \cos(p_1\wedge p_2),\\
G A B \quad  {\mbox {vertex:}} && \quad 2 ig \cos (p_1\wedge p_2),\\
\overline \psi \psi A\quad {\mbox {vertex:}} &&\quad -ig 
\cos (p_1\wedge p_2),\\ 
\overline \psi \psi B\quad {\mbox {vertex:}} &&\quad -ig\gamma_5 
\cos (p_1\wedge p_2).
\end{eqnarray}
\end{mathletters}

Due to the oscillating factors provided by the cosines some of the
integrals constructed with the above rules will be finite but in general
divergences will survive, the degree of superficial divergence
for a generic 1PI graph $\gamma$ being 

\begin{equation}
d(\gamma)= 4 -  I_{AF} -I_{BF}-N_A-N_B-2 N_F-2N_G - \frac32 N_\psi\label{5},
\end{equation}

\noindent
where $N_{\cal O}$ denotes the number of external lines associated to the 
field ${\cal O}$ and $ I_{AF}$ and $I_{BF}$ are the numbers of internal
lines associated to the indicated mixed propagators. In all cases we
will regularize the divergent Feynman integrals by using the
supersymmetric regularization method proposed in \cite{Iliopoulos}. 

\section{THE ONE LOOP APPROXIMATION}
 
It is straightforward to verify that, at the one loop level, all the
tadpoles contributions add up to zero. This confirms the statement
made in the previous section concerning the validity of the Ward
identities. 

\begin{figure}
\centerline{\epsfbox{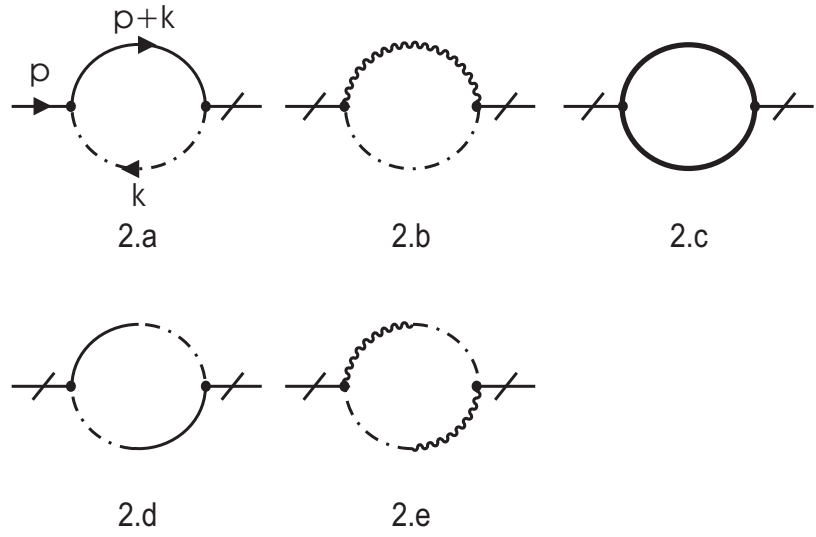}}
\caption{{\it One--loop contributions to the self-energy of the $A$ field.}}
\label{Fig2}
\end{figure}

Let us now examine the contributions to the self-energy of the $A$
field. The corresponding graphs are
those shown in Fig \ref{Fig2}$a-$2$e$. In that figure diagrams $a,b$ and $ c$
are quadratically divergent whereas graphs $d$ and $e$ are
logarithmically divergent.  We shall first prove that the quadratic
divergences are canceled. In fact, we have that

\begin{eqnarray}
\Gamma_{{\mbox 2}a-c}(AA)&=&-g^2\int \frac{d^4k}{(2\pi)^4 }\cos^2(k\wedge p) 
\{4 k^2+4 k^2-2
{\rm Tr}[(\not \!k+\not \!p+m)(\not \!k+m)]\}\nonumber \\
&\phantom a&\times  \Delta(k+p)  \Delta(k),\label{6}
\end{eqnarray}

\noindent
where the terms in curly brackets correspond to the graphs $a$, $b$ and $c$,
respectively. After calculating the trace we obtain

\begin{equation}
\Gamma_{{\mbox 2}a-c}(AA)=8 g^2\int \frac{d^4k}{(2\pi)^4 }(p\cdot k + m^2)
\cos^2(k\wedge p) \Delta(k) \Delta(k+p).\label{7}
\end{equation}

\noindent
This last integral is, at most, linearly divergent. However, the
would be linearly divergent term vanishes by symmetric integration
thus leaving us with an integral which is, at most, logarithmically
divergent.  Adding to Eq.(\ref{7}) the contribution of the graphs 2$d$
and 2$e$ one arrives at 

\begin{equation}
\Gamma_{{\mbox 2}a-e}(AA)= 8 g^2\int \frac{d^4k}{(2\pi)^4 }
\cos^2(p\wedge k) (p\cdot k)\label{8} \Delta(k) \Delta(k+p).
\end{equation}

\noindent 
To isolate the divergent contribution to $ \Gamma_{{\mbox 2}a-e}(AA)$
we Taylor expand the coefficient of $\cos^2(p\wedge k)$ with respect
to the variable $p$ around $p=0$, namely,

\begin{eqnarray}
&& 8 g^2\int\frac{d^4k}{(2\pi)^4}\cos^2(p\wedge k) t^{(1)}(p)\left[ (p\cdot k) \Delta(k) \Delta(k+p)\right]\Bigr.|_{p=0}\nonumber \\
&=& 16 g^2\int\frac{d^4k}{(2\pi)^4}\cos^2(p\wedge k)\frac{(p \cdot k)^2}{(k^2- m^2)^3}  ,   \label{81}
\end{eqnarray}

\noindent 
where $t^{(r)}(p)$ denotes the Taylor operator of order $r$. Since 
$\cos^2(k\wedge p)= (1+\cos(2k\wedge p))/2$ the divergent part of
(\ref{81}) is found to read

\begin{equation}
\Gamma_{Div}(AA)= 2 g^2 p^2 \int \frac{d^4k}{(2\pi)^4 } \frac{1}{(k^2-m^2)^2}\equiv
i I_\xi g^2 p^2\label{9},
\end{equation}

\noindent
where the subscript $\xi$ remind  us that all  integrals
are regularized through the procedure indicated in \cite{Iliopoulos}.
In the commutative Wess-Zumino model this divergence occurs with a
weight twice of the above. As usual, it is 
eliminated by the wave function renormalization $A=Z^{1/2}A_r$,
where $A_r$ denotes the renormalized $A$ field. Indeed, it is easily
checked that with the choice $ Z=1- I_\xi g^2$ the contribution
(\ref{9}) is canceled.

We turn next into analyzing the term containing $\cos(2k\wedge p)$ in
(\ref{81}). For small values of $p$ it behaves as $p^2 \ln (p^2/m^2)$. Thus, in
contradistinction to the nonsupersymmetric $\phi^4_4$ case
\cite{Minwalla}, there is no infrared pole and the function actually
vanishes at $p=0$.

One may  check that at one-loop the $B$ field self-energy is the
same as the self-energy for the $A$ field, i. e., $\Gamma(BB)=\Gamma(AA)$.
Therefore the divergent part of $\Gamma(BB)$ will be eliminated if
we perform the same wave function renormalization as we did for the
$A$ field, $B= Z^{1/2}B_r$. We also found that the mixed two point Green
functions do not have one-loop radiative corrections, $\Gamma(AF)=\Gamma(BG)=0$.

\begin{figure}
\centerline{\epsfbox{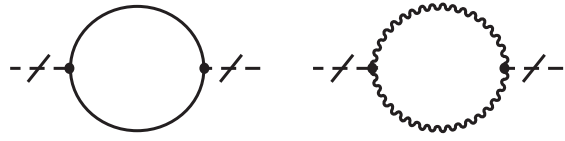}}
\caption{{\it One--loop corrections to the two point function of the auxiliary field $F$. }}
\label{Fig3}
\end{figure}

The one-loop corrections to the two point of the auxiliary field $F$
are depicted in Fig \ref{Fig3}. The two graphs give identical contributions
leading to the result

\begin{equation}
\Gamma(FF)= -4 g^2 \int \frac{d^4k}{(2\pi)^4 }\cos^2(k\wedge p) \Delta(k)
\Delta(k+p),\label{10}
\end{equation}

\noindent 
whose divergent part is
\begin{equation}
\Gamma_{Div}(FF)= 2 g^2 \int    \frac{d^4k}{(2\pi)^4 }\frac{1}{(k^2-m^2)^2}=
i I_\xi g^2,\label{11}
\end{equation}

\noindent
involving the same divergent integral of the two point functions of
the basic fields. It can be controlled by the field renormalization
$F= Z^{1/2} F_r$, as in the case of $A$ and $B$. Analogous reasoning
applied to the auxiliary field $G$ leads to the conclusion that $G
=Z^{1/2}G_r$.  However, things are different as far as the term
containing $\cos(2 k \wedge p)$ is concerned. It diverges as
$\ln(p^2/m^2)$ as $p$ goes to zero. Nevertheless, this is a harmless
singularity in the sense that its multiple insertions in higher order
diagrams do not produce the difficulties pointed out in
\cite{Minwalla}.

\begin{figure}
\centerline{\epsfbox{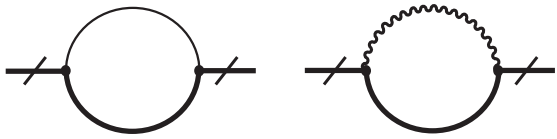}}
\caption{{\it One--loop contributions to the self-energy of the spinor field
$\psi$.}}
\label{Fig4}
\end{figure}

Let us now consider the corrections to the self-energy of the spinor field
$\psi$ which are shown in Fig. \ref{Fig4}. The two contributing graphs give

\begin{eqnarray}
\Gamma(\psi \overline \psi)&=& 4 g^2 \int \frac{d^4 k}{(2\pi)^4}
\cos^2(k\wedge p) \Delta(k)\Delta(k+p)[(\not \! k +m)- \gamma_5(\not\! k+m)\gamma_5]
\nonumber\\
& =& 8 g^2 \int \frac{d^4 k}{(2\pi)^4}
\cos^2(k\wedge p) \not \! k\, \Delta(k)\Delta(k+p),\label{12}
\end{eqnarray}

\noindent
so that for the divergent part we get $\Gamma_{Div}(\psi\overline
\psi)= ig^2\not\!\! p \,I_\xi$ leading to the conclusion that the
spinor field presents the same wave function renormalization of the
bosonic fields, i. e., $\psi = Z^{1/2} \psi_r$.  As for the term
containing $\cos(2k\wedge p)$ it behaves as $\not \! p
\ln(p^2/m^2)$ and therefore vanishes as $p$ goes to zero.

\begin{figure}
\centerline{\epsfbox{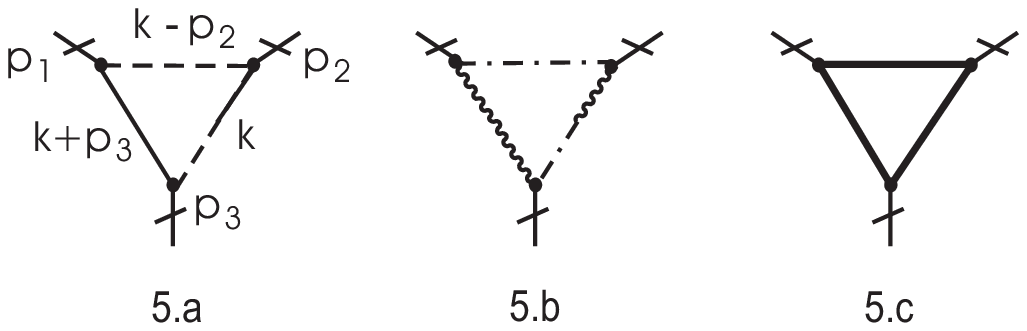}}
\caption{{\it Divergent graphs contributing to the three point function  of the $A$ field.}}
\label{Fig5}
\end{figure}

The one-loop superficially (logarithmically) divergent graphs
contributing to the three point function of the $A$ field are shown in
Fig \ref{Fig5}. The sum of the amplitudes corresponding to the graphs
\ref{Fig5}$a$ and \ref{Fig5}$b$ is

\begin{eqnarray}
\Gamma_{5a+5b}(AAA)&=& 96 ig^3 m \int\frac{d^4k}{(2\pi)^4} (k-p_2)^2 \Delta(k) \Delta(k+p_3)
\Delta(k-p_2)\cos( k\wedge p_1 + p_3\wedge p_1) \nonumber\\
&\phantom a &\times\cos(p_2\wedge k)\cos(p_3\wedge k),\label{13}
\end{eqnarray}

\noindent
while its divergent part is found to read

\begin{equation}
\Gamma_{5a+5b\,\,Div}(AAA)= 24 ig^3 m \cos( p_3\wedge p_1) \int
\frac{d^4k}{(2\pi)^4} (k)^2 (\Delta(k))^3.
\end{equation}

\noindent
The divergent part of the graph 5$c$, nonetheless, gives a similar
contribution but with a minus sign so that the two divergent parts add
up to zero. Thus, up to one-loop the three point function
$\Gamma(AAA)$ turns out to be finite. Notice that a nonvanishing
result would spoil the renormalizability of
the model. The analysis of $\Gamma(ABB)$ follows along similar lines
and with identical conclusions. Furthermore, it is not difficult to
convince oneself that $\Gamma(FAA)$, $\Gamma(FBB)$ and $\Gamma(GAB)$
are indeed finite. 

\begin{figure}
\centerline{\epsfbox{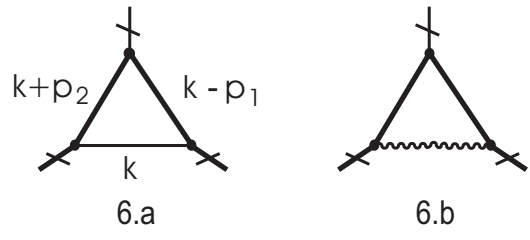}}
\caption{{\it One--loop contributions to the three point function $\Gamma(A\overline \psi\psi)$}}
\label{Fig6}
\end{figure}

As for $\Gamma(A \psi\overline \psi)$ we notice that superficially
divergent contributions arise from the diagrams depicted in Figs
\ref{Fig6}$a$ and \ref{Fig6}$b$. In particular, diagram \ref{Fig6}$a$
yields

\begin{eqnarray}
\Gamma_{6a}(A\psi\overline\psi)&=&8ig^3\int \frac{d^4k}{(2\pi)^4} \Delta(k) 
\Delta(p_2+k) \Delta (k-p_1)
(\not \! p_2+\not \! k+m)(\not \! k-\not \! p_1+m)\nonumber\\
&\phantom a&\times \cos(k\wedge p_3-p_1\wedge p_3) \cos(k\wedge p_1)\cos(k\wedge p_2), \label{14}
\end{eqnarray}

\noindent
while 6$b$ gives

\begin{eqnarray}
\Gamma_{6b}(A\psi\overline\psi)&=&-8ig^3\int \frac{d^4k}{(2\pi)^4} \Delta(k) 
\Delta(p_2+k) \Delta( k-p_1)\gamma_5(\not \! p_2+\not \! k+m)
(\not \! k-\not \! p_1+m)\gamma_5\nonumber \\
&\phantom a&\times \cos(k\wedge p_3-p_1\wedge p_3) \cos(k\wedge p_1)
\cos(k\wedge p_2) , \label{15}
\end{eqnarray}

\noindent
so that the sum of the two contributions is also finite. The same
applies for $\Gamma(B\psi\overline\psi)$. 

We therefore arrive at another important result, namely, that there is
no vertex renormalization at the one loop level. This parallels the
result of the commutative Wess-Zumino model. 

\begin{figure}
\centerline{\epsfbox{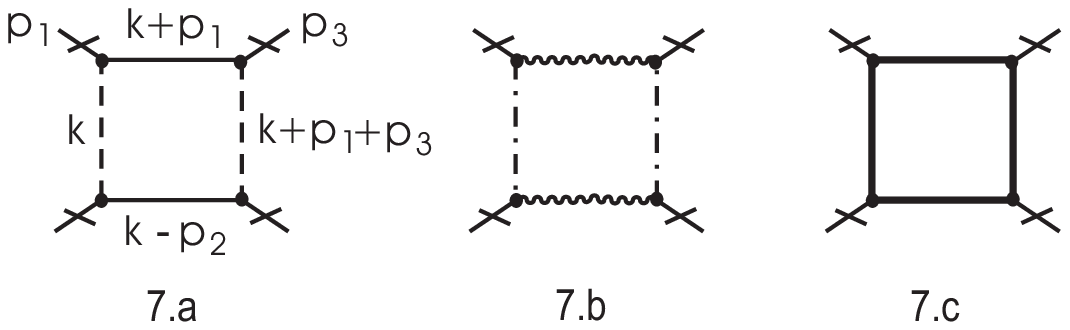}}
\caption{{\it Divergent graphs contributing to the four point function of the
$A$ field.}}
\label{Fig7} 
\end{figure}

To complete the one-loop analysis we must examine the four point
functions. Some of the divergent diagrams contributing to
$\Gamma(AAAA)$ are depicted in Fig. \ref{Fig7}$a-c$.  The analytical expression
associated with the graph \ref{Fig7}$a$ is

\begin{eqnarray}
\Gamma_{7a}(AAAA) &=& 16g^4\int \frac{d^4k}{(2\pi)^4}k^2\Delta(k)\Delta (k+p_1)
(k+p_1+p_3)^2 \Delta(k+p_1+p_3)\Delta (p_2 -k)\nonumber \\
&\phantom a & \times\cos(k\wedge p_1) \cos(k\wedge p_2) 
\cos [(k+p_1)\wedge p_3]
\cos[(k-p_2)\wedge p_4].
\label{16}
\end{eqnarray} 

\noindent
There are five more diagrams of this type, which are obtained by
permuting the external momenta $p_2\,\, , p_3$ and $p_4$ while keeping
$p_1$ fixed. Since we are interested in the (logarithmic) divergence
associated with this diagram, we set all the external momenta
to zero in the propagators but not in the arguments of the cosines. This 
yields

\begin{eqnarray}
\Gamma_{7a\,\, Div}(AAAA) &=& 16g^4\int \frac{d^4k}{(2\pi)^4}(k^2)^2
(\Delta(k))^4\nonumber \\
&\phantom a & \times
\cos(k\wedge p_1) \cos(k\wedge p_2) \cos [(k+p_1)\wedge p_3]
\cos[(k-p_2)\wedge p_4].\label{17}
\end{eqnarray}

\noindent
Adopting the same procedure for the other five graphs we notice 
that the corresponding contributions are pairwise equal. The final
result is therefore

\begin{eqnarray}
&&\Gamma_{Div}(AAAA) = 32g^4\int \frac{d^4k}{(2\pi)^4}(k^2)^2
(\Delta(k))^4  \cos(k\wedge p_1)\nonumber \\
&\phantom a &\times [ \cos(k\wedge p_2) \cos [(k+p_1)\wedge p_3]
\cos[(k-p_2)\wedge p_4] +p_3\leftrightarrow p_4+p_2\leftrightarrow p_4].
\label{18}
\end{eqnarray}

There is another group of six diagrams, Fig \ref{Fig7}$b$, which are obtained 
from the preceding
ones by replacing the propagators of $A$ and $F$ fields by the propagator of 
the $B$ and $G$ fields, respectively. The net effect of adding these
contributions is, therefore, just to double the numerical factor in the right
hand side of the above formula.

Besides the two groups of graphs just mentioned, there are another six
graphs with internal fermionic lines. A representative of this group
has been drawn in Fig \ref{Fig7}$c$. It is straightforward to verify that
because of the additional minus sign due to the fermionic loop, there
is a complete cancellation with the other contributions described
previously.  The other four point functions may be analyzed similarly
with the same result that no quartic counterterms are needed.

\section{Absence of Mass  and Coupling Constant Renormalization to all
  Orders of Perturbation Theory}

In the previous section we proved that up to one loop the
noncommutative Wess-Zumino model is renormalizable and only requires a
common wave function renormalization. Here, we shall
prove that no mass and coupling constant counterterms are needed at
any finite order of perturbation theory. As in the commutative case,
our proof relies heavily on the Ward identities. 

We start by noticing that from Eq.(\ref{1}) it follows that 

\begin{equation}
\int d^4 y\frac{\delta}{\delta {\cal O}(y)}\int d^4 x\,
\underbrace{{\cal O}(x)\star{\cal O}(x)\star\ldots \star{\cal O}(x)}_{n \rm 
\; factors}= n \int d^4 x\,\underbrace{{\cal O}(x)\star{\cal O}(x)\star
\ldots \star{\cal O}(x)}_{n-1 \rm \; factors}. \label{21}
\end{equation}

\noindent
In turns, this enables one to find 

\begin{equation}
\frac{\partial \phantom a}{\partial m} Z(J)= -\frac{m}{2 g}\int \frac{\delta
Z(J)}{\delta J_F(y)} d^4 y - \frac{iZ(J)}{2g} \int J_A(y) d^4 y, \label{19}
\end{equation} 

\noindent
which looks formally identical to the corresponding relation in the
commutative case \cite{Iliopoulos}.  Here, $Z(J)$ is the Green
function generating functional and $J_{\cal 
O}$ is the external source associated to the field ${\cal
O}$. By collectively denoting the fields by  $\phi$, $Z(J)$ can
be cast as

\begin{equation}
Z(J)= \int D\phi \exp i \left(S+ \int d^4 x J \phi\right), \label{20}
\end{equation}

\noindent 
where $S=\int d^4x {\cal L}$ and ${\cal L}$ is the regularized Lagrangian. 

In terms of the 1PI generating functional $\Gamma(R)$ the identity
(\ref{19}) becomes 

\begin{equation}
\frac{\partial}{\partial m}\Gamma[R]= - \frac{m}{2g}\int R_F(y)d^4 y + 
\frac{1}{2g}\int\frac{\delta\Gamma[R]}{\delta R_A(y)} d^4 y.\label{22} 
\end{equation}

\noindent 
By taking the functional derivative with respect to $R_F$ and then
putting all $R$'s equal to zero we obtain

\begin{equation}
m= \Gamma(FA)\Bigl |_{p^2=0}= Z^{-1}\Gamma_r(FA) \Bigl |_{p^2=0},  \label{23}
\end{equation}

\noindent
where $\Gamma_r(AF)$ is the renormalized 1PI Green function of the
indicated fields. We take as normalization conditions those specified
in \cite{Iliopoulos}. Specifically, $\Gamma_r(FA) \Bigl
|_{p^2=0}=m_r$, where $m_r$ is taken to be the renormalized
mass. Hence, $m_r=Z m$ implying that there is no additive mass
renormalization. Through similar steps one also finds that
$g_r=Z^{3/2}g$, where $g_r$ is the renormalized coupling
constant. This implies the absence of coupling constant counterterms.

We stress the fact that, by exploiting the Ward identities, we have
succeeded in generalizing to all orders of perturbation theory the one
loop result concerned with the absence of counterterms different from those
already present in the original Lagrangian.

\section{Conclusions}

After extending the Wess-Zumino model to the noncommutative Minkowski
space, we succeeded in demonstrating, to all orders of perturbation,
that the theory is free of nonintegrable infrared singularities
and renormalizable. Thus, this model provides an example of a fully
consistent noncommutative quantum field theory. 

It shares some properties with the Wess-Zumino model. The
quadratic and linear divergences are absent. Furthermore, only a
wave function renormalization is needed to make the theory finite.
Also, all fields exhibit the same mass as is the case in any ordinary
supersymmetric theory.    

On the other hand, one should notice that the commutative Wess-Zumino
model can not be recovered from the noncommutative one at the limit
of vanishing deformation. In fact, the limit of vanishing deformation
does not exist because of  logarithmic singularities. 

A very important feature of the noncommutative Wess-Zumino model is that
all vertices were deformed in the same way. This was essential to split
the amplitudes into planar and non planar contributions in a uniform way
so that the renormalizability properties of the Wess-Zumino model is
always present in the planar sector. The reason for the deformed vertices
to be the same is the presence of the auxiliary fields. With them all
interactions are cubic. The elimination of the auxiliary fields produces
cubic and quartic interactions and the vertices will be deformed in different
ways. Of course, the renormalizability properties will be the same without
the auxiliary fields but, surely, more difficult to prove. Supersymmetric
gauge theories have cubic and quartic vertices even in the presence
of auxiliary fields. We expect that the renormalizability proof will be much
more difficult unless further simplification arise. Studies in this direction
are in progress.

\begin{center}
{ACKNOWLEDGMENTS}
\end{center}

 This work was partially supported by Funda\c c\~ao de Amparo
\`a Pesquisa do Estado de S\~ao Paulo (FAPESP) and Conselho Nacional de
Desenvolvimento Cient\'\i fico e Tecnol\'ogico (CNPq).

 \newpage
 \newpage
 \newpage
 \newpage
 \newpage
 \newpage
\end{document}